%% file: main.tex
\documentclass{article}

\PassOptionsToPackage{numbers, compress}{natbib}

\usepackage[preprint]{neurips_2022}

\usepackage[utf8]{inputenc} 
\usepackage[T1]{fontenc}    
\usepackage{hyperref}       
\usepackage{url}            
\usepackage{booktabs}       
\usepackage{amsfonts}       
\usepackage{nicefrac}       
\usepackage{microtype}      
\usepackage{xcolor}         

\usepackage{amsmath}
\usepackage{amssymb}
\usepackage{mathtools}
\usepackage{amsthm}

\usepackage[capitalize,noabbrev]{cleveref}

\usepackage{microtype}
\usepackage{graphicx}
\usepackage{subfigure}
\usepackage{booktabs} 
\usepackage{multirow}
\usepackage{array}
\usepackage{braket}

\usepackage{wrapfig}

\usepackage{Algorithm}
\usepackage{Algorithmic}

\title{Lattice Convolutional Networks for Learning Ground States of Quantum Many-Body Systems}

\author{%
  Cong Fu\thanks{Equal contribution} \\
  Computer Science and Engineering\\
  Texas A\&M University\\
  College Station, TX 77843, USA \\
  \texttt{congfu@tamu.edu} \\
   \And
   Xuan Zhang\footnotemark[1] \\
  Computer Science and Engineering\\
  Texas A\&M University\\
  College Station, TX 77843, USA \\
   \texttt{xuan.zhang@tamu.edu} \\
  \And
   Huixin Zhang\\
  Computer Science and Engineering\\
  Texas A\&M University\\
  College Station, TX 77843, USA \\
   \texttt{zhanghui21@tamu.edu} \\
   \And
   Hongyi Ling\\
  Computer Science and Engineering\\
  Texas A\&M University\\
  College Station, TX 77843, USA \\
   \texttt{hongyiling@tamu.edu} \\
   \And
   Shenglong Xu\thanks{Corresponding authors}\\
  Physics and Astronomy\\
  Texas A\&M University\\
  College Station, TX 77843, USA \\
   \texttt{slxu@tamu.edu} \\
   \And
   Shuiwang Ji\footnotemark[2]\\
  Computer Science and Engineering\\
  Texas A\&M University\\
  College Station, TX 77843, USA \\
   \texttt{sji@tamu.edu} \\
}

\begin{document}

\maketitle

\begin{abstract}
Deep learning methods have been shown to be effective in representing ground-state wave functions of quantum many-body systems. Existing methods use convolutional neural networks (CNNs) for square lattices due to their image-like structures. For non-square lattices, existing method uses graph neural network (GNN) in which structure information is not precisely captured, thereby requiring additional hand-crafted sublattice encoding. In this work, we propose lattice convolutions in which a set of proposed operations are used to convert non-square lattices into grid-like augmented lattices on which regular convolution can be applied. Based on the proposed lattice convolutions, we design lattice convolutional networks (LCN) that use self-gating and attention mechanisms. Experimental results show that our method achieves performance on par or better than existing methods on spin 1/2 $J_1$-$J_2$ Heisenberg model over the square, honeycomb, triangular, and kagome lattices while without using hand-crafted encoding.
\end{abstract}

\section{Introduction}
Study of quantum many-body problems is of fundamental interests in physics. It is crucial for theoretical modeling and simulation of complex quantum systems, materials and molecules~\cite{carleo2019machine}. For instance, graphene, arguably the most famous 2D material, is made of carbon atoms on a honeycomb lattice. Solving quantum many-body problems remains to be very challenging because of the exponential growth of Hilbert space dimensions with the number of particles in quantum systems. Only approximation solutions are available in most cases. Tensor network~\cite{white1992density,schollwock2011density,orus2014practical,biamonte2017tensor} is one of the popular techniques to model quantum many-body systems but suffers entanglement problems~\cite{choo2018symmetries}. Variational Monte Carlo (VMC)~\cite{mcmillan1965ground} is a more general methodology to obtain quantum many-body wave functions by optimizing a compact parameterized variational ansatz with data sampled from itself. But how to design variational ansatz with high expressivity to represent real quantum states is still an open problem.

Recently traditional machine learning models, such as restricted Boltzmann machine (RBM)~\cite{smolensky1986information}, has been used as variational ansatz~\cite{carleo2017solving,nomura2017restricted,choo2018symmetries,kaubruegger2018chiral,choo2020fermionic,nomura2021helping}. Following this direction, some studies explore deep Boltzmann machines~\cite{gao2017efficient,carleo2018constructing,pastori2019generalized} and fully-connected neural networks to represent quantum states~\cite{saito2018machine,cai2018approximating,saito2017solving,saito2018method,saito2018machine}. Most recent studies also use CNN as variational ansatz for square lattice systems~\cite{liang2018solving,choo2019two,zheng2021speeding,liang2021hybrid,roth2021group}. And GNN has been applied to non-square lattices and random graph systems~\cite{yang2020scalable,kochkov2021learning}. 

In this work, we explore the potential of using CNN as variational anstaz for non-square lattice quantum spin systems. We propose lattice convolutions that use a set of proposed operations to convert non-square lattices into grid-like augmented lattices on which any existing CNN architectures can be applied. Based on proposed lattice convolution, we  design highly expressive lattice convolutional networks (LCN) by leveraging self-gating and attention mechanisms. Experimental results show that our methods achieve performance on par or better than state-of-the-art methods over the square, honeycomb, triangular, and kagome lattice quantum systems on spin 1/2 $J_1$-$J_2$ Heisenberg model, a prototypical quantum many-body model of magnetic materials that captures the exchange interaction between spins. 

\textbf{Relations with Prior Work. }LCN uses same experiment setting and training strategy as in~\citet{kochkov2021learning}, where hand-crafted sublattice encoding is needed as extra input node feature to a GNN model to explicitly encode the lattice structure information. However, LCN only need to use easily sampled spin configurations as network input. There is also another work on surface lattices~\cite{lym2019lattice} that requires additional symmetry permutations in convolution. Further, convolutions on irregular shapes are also developed in computer vision~\cite{hanocka2019meshcnn,hu2022subdivision,su2018splatnet}, but are not suitable for quantum systems.

\section{Background and Related Work}
In quantum mechanics, a quantum state is represented as a vector in Hilbert space. This vector is a linear combination of observable system configurations $\{c_i\}$, known as a computational basis. In the context of spin 1/2 systems, each spin can be measured in two states, spin-up or spin-down, which are represented by $\uparrow$ and $\downarrow$, respectively. All the combinations of spins form a set of basis. Given $N$ spins, there are in total $2^N$ configurations in the computational basis. Specifically, a state can be written as
\begin{align} \label{eq:state}
    |\psi\rangle = \sum_i^{2^N} \psi(c_i) |c_i\rangle,
\end{align}
where $\ket{c_i}$ represents an array of spin configurations of $N$ spins, e.g., $\uparrow \uparrow \downarrow \cdots \downarrow$, and $\psi(c_i)$ is the wave function, which is in general a complex number.  The summation is over all possible $2^N$ spin configurations. 
The squared norm $|\psi(c_i)|^2$ corresponds to the probability of system collapsing to configuration $c_i$ when being measured, and $\sum_i^{2^N}|\psi(c_i)|^2=1$ due to normalization.

\subsection{Ground States}
The ground state of a quantum system is its lowest-energy state. Usually, many physical properties can be determined by the ground state. Particle interactions within a given quantum many-body system are determined by a Hamiltonian, which is an Hermitian matrix $\mathbf{H}$ in the Hilbert space. System energy and its corresponding quantum state are governed by the time-independent Schrödinger equation:
\begin{align}\label{eq:schrodinger}
    \mathbf{H} |\psi\rangle = E |\psi\rangle,
\end{align}
which is an eigenvalue equation. The eigenenergy $E$ is the eigenvalue of $\mathbf{H}$ and $|\psi\rangle$ is the corresponding eigenvector. In principle, those can be obtained by eigenvalue decomposition given $\mathbf{H}$. The lowest eigenvalue is called the ground state energy, and its associated eigenvector is called the ground state. The ground state and the ground state energy determine the property of the quantum system at zero temperature.

\subsection{Variational Principle in Quantum Mechanics}
\label{section:varional learning}
Given a system of size $N$, the dimension of the Hamiltonian matrix is $2^N \times 2^N$. Since the dimension of the matrix grows exponentially with system size, it is intractable to use eigenvalue decomposition directly even for relatively small systems. The state-of-the-art algorithm using Lanczos method, that explores the sparseness of $\mathbf{H}$, can obtain the ground state energy and the ground state for $N$ up to $\sim 30$.
For larger systems, a common approach is to use variational principle to approximately solve the Schrödinger equation. According to the variational principle, the energy of any given quantum state is greater than or equal to the ground state energy. So we can optimize parameterized wave functions to make the energy as low as possible. Specifically, we can approximate the ground state of a Hamiltonian $\mathbf{H}$ by minimizing the variational term $E$ shown as below:  
\begin{align}
    E = \frac{\langle\psi|\mathbf{H}|\psi\rangle}{\langle\psi|\psi\rangle} \ge E_0,
\end{align}
where $E$ is the expectation value of energy of a variational quantum state $|\psi\rangle$ for a given Hamiltonian $\mathbf{H}$, and $E_0$ is the true ground state energy. The state $|\psi\rangle$ takes the form of Eq.~\eqref{eq:state}. Given $\mathbf{H}$, the expectation value $E$ is determined by the wavefunction $\psi(c_i)$ that
can be any parameterized functions. The goal is to find the optimal function $\psi(c_i)$ that minimizes $E$. The success of the variational method relies on the expressivity of the parameterized function. Therefore it is natural to explore neural networks as variational ansatz of the wavefunction.

\subsection{Related Work}
Variational quantum many-body states with wave functions given by neural networks are called neural-network quantum states, initially studied by~\citet{carleo2017solving}, where they use RBM to represent many-body wave function. Subsequent studies~\cite{choo2018symmetries,saito2018method,cai2018approximating} apply fully-connected neural networks as variational ansatz, which has been shown to be more effective than RBM methods. But these methods do not explicitly consider the structure information when applied to two-dimentional systems~\cite{cai2018approximating}. Motivated by the spatial symmetry of periodic quantum systems and successful practices of convolutional neural networks (CNNs) in computer vision~\cite{krizhevsky2012imagenet}, \citet{liang2018solving}, \citet{choo2019two} and \citet{szabo2020neural} use CNNs to represent quantum states for square lattices. CNN is able to effectively represent highly entangled quantum systems~\cite{levine2019quantum} than RBM based representations~\cite{deng2017quantum}, which benefits from its information reuse. However, CNN cannot be naturally used on non-grid like systems. Recently graph neural networks (GNNs) have also been applied to represent wave functions~\cite{yang2020scalable,kochkov2021learning}. GNNs can work with arbitrary geometric lattices or even random graph systems but structure information is not precisely captured. So additional hand-crafted sublattice encoding is needed to augment system configurations in order to respect underlying quantum symmetry~\cite{kochkov2021learning}. 

Wave functions are usually complex-valued, therefore it is necessary to predict both amplitudes and phases.~\citet{choo2019two} use complex-valued weights and biases to predict amplitudes and phases simultaneously, and design a generalized activation function. Whereas amplitudes and phases are predicted separately using real-valued networks in~\citet{szabo2020neural,kochkov2021learning}.

\section{Lattice Convolutional Networks}

While GNN can be naturally applied to non-square lattices, isotropic weights prevent it from capturing rich structure information. Therefore auxiliary hand-crafted structure encoding is needed to augment the original spin configuration input on lattices. We argue that CNN is more suitable to model wave functions for lattice systems, which features repetitive local patterns. 

In this section, we introduce LCN, a novel lattice convolutional network that has strong capability to model wave functions for non-square lattice systems without using any extra structure encoding. 

\subsection{Motivation and Overview}
\label{sec:overview}
\begin{figure*}[t]
\begin{center}
\centerline{\includegraphics[width=0.9\textwidth]{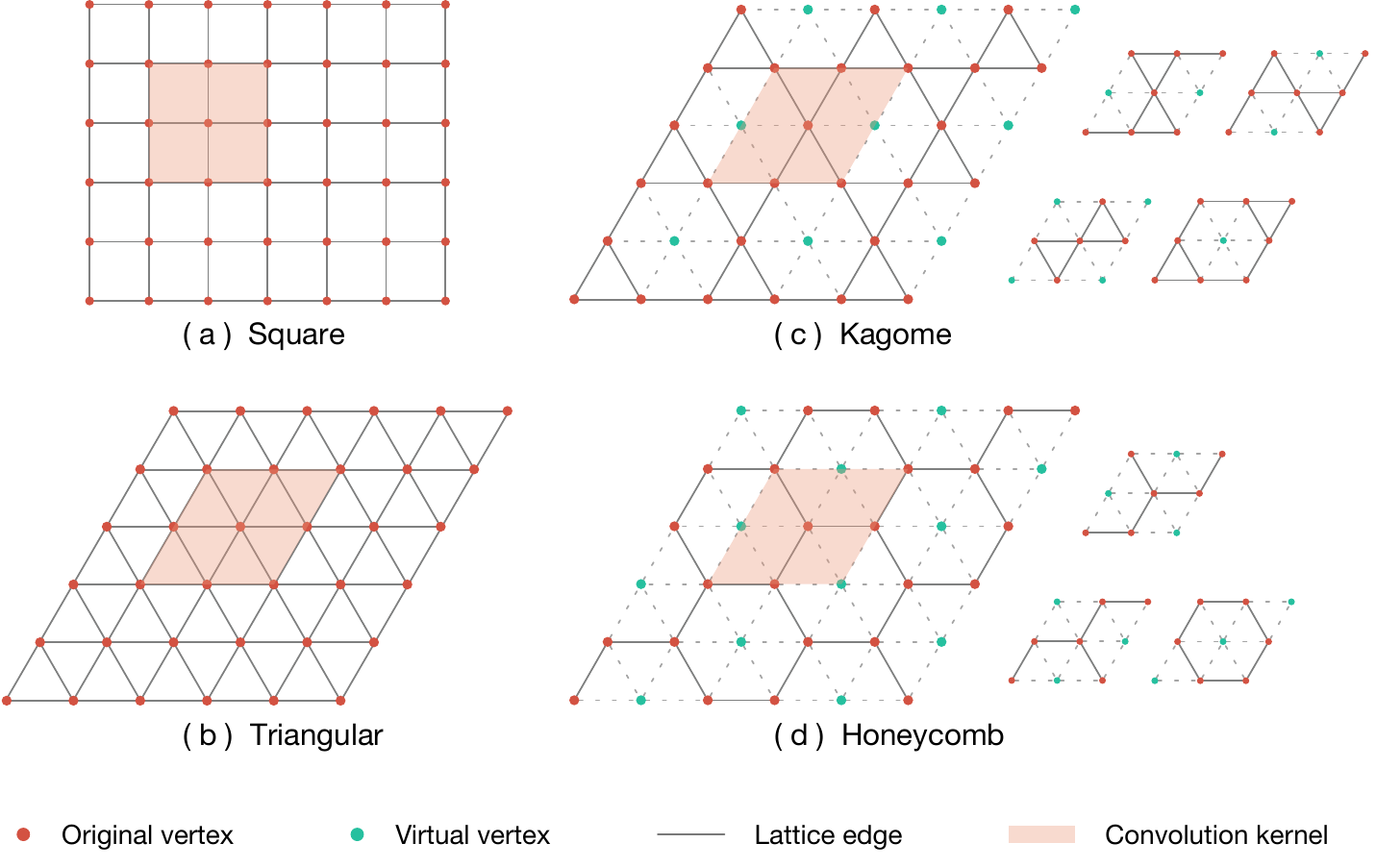}}
\vspace{-0.1in}
\caption{Structures of four different lattices studied in this work. Shadow areas denote the regular convolution kernels. Red dots represent original vertices in lattices, and green dots represent padded virtual vertices. In (c) and (d), dashed lines denote the nearest neighbors of virtual vertices. On the right are different substructures captured by convolution. Note that shear directions won't affect the result due to lattice symmetry. Even though the lattices shown are finite but they are periodically arranged in whole space, which is realized by applying periodic boundary condition.}
\label{fig:lattice_kernel}
\end{center}
\vskip -0.4in
\end{figure*}

In this work, we focus on designing CNNs on four types of lattice, including square, triangular, honeycomb and kagome, which are the four of the most common lattice structures that describes two dimensional materials. The lattice structures are shown in \cref{fig:lattice_kernel}.
CNNs are known to be efficient feature extractors on regular structures. Inside the network, each convolution layer applies anisotropic filters on local regions, shared across the entire input. Through training, each filter learns to be sensitive to different local patterns. By stacking multiple convolution layers, the global structure information can be precisely captured.

It is straightforward to apply CNNs on a square lattice due to its image-like structure. Triangular lattices can be viewed as sheared square lattices where every unit cell of the square lattice undergoes the same affine transformation. Therefore we can shear the convolution kernel accordingly to match the shape of the transformed unit cells. However, for lattices that cannot be converted into grids, such as honeycomb and kagome, the key is to define the shape of the convolution kernel and optimize the weight sharing across convolution sites. In the proposed LCN, we solve these challenges in a principled way by converting non-square lattices into grid-like augmented lattices through a set of operations such that regular convolution kernels can be applied.

\subsection{Augmented Lattices}
\label{sec:Augmented lattice}

As mentioned in~\cref{sec:overview}, triangular lattices can be seen as sheared square lattices, which implies that the local structure is same everywhere on the lattice, hence regular square kernels can be naturally applied. By contrast, honeycomb and kagome lattices have multiple local structures, making it difficult to share kernel weights across different structures. Moreover, different local structures are arranged in a staggered manner, which impedes the information reuse among the same local structures. 

Critically, we make the key observation that honeycomb and kagome lattices can be converted from triangular lattices by removing some vertices and edges. Conversely, honeycomb and kagome lattices can be viewed as triangular lattices through augmenting virtual vertices back on the original lattices. As shown in \cref{fig:lattice_kernel} (c) and (d), virtual vertices (green dots) are inserted in the center of each hexagon sub-structure. Through this augmenting operation, we can apply identical kernel everywhere on the lattices regardless of original local structures but still can capture different structure information.

The advantages of this operations are two folds. First, during convolution, the virtual vertices participate in the convolution in the same way as the original vertices. i.e., values of virtual vertices are also updated. By doing so, the virtual lattices can gather and distribute the information from the original vertices, which can help increase receptive field and boost information exchange. Second, by adding virtual vertices, we can overlap the same convolution kernel in order to enable information reuse, which is crucial to capture long-range spin correlations~\cite{liang2018solving}. To some extent, augmenting with virtual vertices could enhance the wave function representation ability.

%



\subsection{Lattice Convolutions}
\label{sec:lattice_conv}
After augmentation, the lattice becomes either square or triangular, both of which are grid-structured. However, due to the characteristics of the input, additional processing steps are required in order to apply regular convolutions on the augmented lattice.

\begin{wrapfigure}{r}{0.5\textwidth}
\begin{center}
\includegraphics[width=0.5\textwidth]{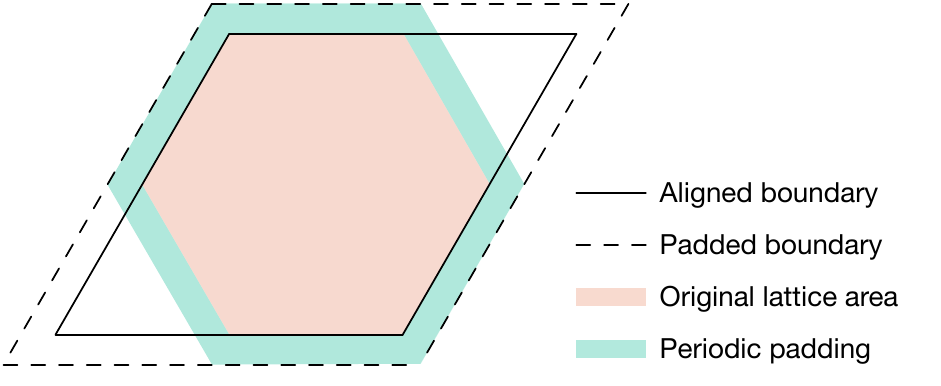}
\caption{Boundary alignment and periodic padding.}
\label{fig:periodic_padding}
\end{center}
\vskip -0.3in
\end{wrapfigure}

\textbf{Boundary alignment (BA). }
Lattices with finite vertices often have irregular boundaries. However, in current deep learning libraries, convolutions on image-like data typically requires the input grid to have equal number of elements on each row. Therefore, we zero-pad the augmented lattices into parallelograms. In our experiments, square lattices already have regular boundaries so this step is omitted. The aligned boundary is drawn with solid line in \cref{fig:periodic_padding}.

\textbf{Periodic padding (PP). }
For images, boundaries are commonly zero-padded before convolutions to preserve the size of feature maps. Finite quantum systems often consider periodic boundary conditions so that the lattice can be repeated to fill the entire space. To preserve this important structure information, after padding the aligned boundaries with zero, we replace the padding values around the original lattice area with the values given by the periodic boundary condition. We can optionally do the periodic padding for the virtual vertices as well. In \cref{fig:periodic_padding}, the padded boundary is drawn with dashed line, the original lattice area is marked with pink shadow and the periodic padding is marked with green shadow.

\textbf{Mask. }
Finally, after each convolution, to clean up the artifacts introduced in the two previous steps, we reset all vertices used for the boundary alignment and the periodic padding to zero. We do not reset the virtual vertices to allow information passing through them, i.e., we only reset the vertices outside the original lattice area.

To summarize, the proposed lattice convolution applied on an input lattice $\mathbf{U}$ is defined as:
\begin{align}
\label{eq:lattice_conv}
\text{LatticeConv}(\mathbf{U};\mathbf{W}) = \textbf{Mask}(\mathbf{W}*\textbf{PP}(\textbf{BA}(\textbf{Aug}(\mathbf{U})))),
\end{align}
where $\mathbf{W}$ is the convolution weight matrix; \textbf{PP}, \textbf{BA} and \textbf{Mask} stand for the above mentioned three processing steps and \textbf{Aug} stands for the augmentation step defined in \cref{sec:Augmented lattice}; The symbol $*$ denotes the regular convolution which is defined as:
\begin{align}
\label{eq:conv}
    (\mathbf{W}*\mathbf{U^\prime})_{i,j} = \sum_{m=-s}^{s}  \sum_{n=-s^\prime}^{s^\prime} \mathbf{W}_{mn} \mathbf{U^\prime}_{i-m, j-n},
\end{align}
where $\mathbf{U}^\prime \in \mathbb{R}^{H \times W \times d}$ denotes a $H\times W$ feature map with $d$ input channels. $\mathbf{W}$ has shape $(2s+1)\times(2s^\prime+1)\times d^\prime \times d$ where $d^\prime$ is the number of output channels and $(2s+1)\times (2s^\prime+1)$ is the size of the receptive field.


\subsection{Instantiations}
\label{sec: Instantiations}

\textbf{Square. }
As shown in \cref{fig:lattice_kernel} (a), for square lattices, the $3 \times 3$ kernel receptive field contains all four nearest neighbors and four second nearest neighbors (or called next nearest neighbor in some references), which is defined by the euclidean distance, instead of by connectivity. 

\textbf{Triangular. }
For triangular lattices, the receptive field of $3 \times 3$ convolution kernel includes six nearest neighbors and two second nearest neighbors of center vertices, as shown in \cref{fig:lattice_kernel} (b). This structure pattern is same at all positions across the lattice.

\textbf{Honeycomb. }
For honeycomb lattices, the $3 \times 3$ convolution kernel centered on the original vertices captures all three nearest neighbors and two second nearest neighbors. As shown in the \cref{fig:lattice_kernel} (d), depending on the local structure, the mapping between the kernel weights and the original vertices can have two different situations. The kernel centered on virtual vertices captures five original vertices around it. For honeycomb lattice, we also apply the periodic padding for the virtual vertices.

\textbf{Kagome. }
For kagome lattices, as shown in \cref{fig:lattice_kernel} (c), the $3 \times 3$ kernel can capture all nearest neighbors and some next nearest neighbors. The virtual vertices can receive information from eight original vertices around it. There are three local structures centered on original vertices. And we only apply the periodic padding for the virtual vertices on kagome lattices of size 36.


\subsection{CNN versus GNN}
\label{sec: CNN vs GNN}
The convolution applies different weights for neighbors with different relative positions, which we argue is critical to capture the structure information. This is to the opposite with graph neural networks where same weights are used for all one-hop neighbors on a graph. For example, we can define a graph on the square lattice by defining edges as the spatially nearest neighbors. For quantum many-body systems, the spatially second nearest neighbors play an important role in defining the energy. But the two-hop neighbors on the graph will also include the third nearest neighbors. To be able to capture the structure information with GNNs,~\citet{kochkov2021learning} propose to augment the vertices with sublattice encoding to explicitly provide structure information at input. Our experiments show that such encoding is indeed critical for GNNs. On the contrary, our lattice convolution accurately learns the ground state without any additional input. To some extent, the proposed lattice convolution can learn the structure encoding automatically in the kernel space.

\section{Network Architecture and Training}
\label{sec:network architecture}
After constructing the grid-like input and defining the convolution operation, any existing convolution neural network architecture can be applied. We aim to design variational ansatz to have powerful expressivity and capture spatial long-range spin correlations. To this end, our model is developed based on recent advanced deep learning modules and attention mechanisms. The details of main components are described below.

\textbf{Squeeze-and-Excitation Block. }
Squeeze-and-Excitation (SE) block~\cite{hu2018squeeze} can improve the quality of representations produced by a network by explicitly modeling the channel-wise interdependencies. It utilizes squeeze operation to aggregate channel-wise global spatial information and excitation operation to capture channel-wise dependencies through self-gating recalibration. Details of formulation can be found in~\cref{app:network}.



\textbf{Non-Local Block. }
We find it is useful to incorporate spin-spin global interaction other than only locality interaction defined by lattice edges. Non-local operation~\cite{wang2018non} is designed to capture long-range dependencies. Specifically, non-local operation make features at one position attend to all other position’s features. Details of formulation can be found in~\cref{app:network}.



\textbf{SE-Non-Local Layer. }
Based on the SE block and non-local block, we propose SE-Non-Local Layer by further introducing skip connection and pre-activation techniques. Therefore, the resulting SE-Non-Local Layer consists of the following components: Normalization $\rightarrow$ Activation $\rightarrow$ LatticeConv $\rightarrow$ SE Block $\rightarrow$ Addition $\rightarrow$ Non-local Block. We apply LayerNorm (LN)~\cite{ba2016layer} as the normalization and ReLU as the activation function.

\begin{figure*}[t]
\vskip 0.2in
\begin{center}
\centerline{\includegraphics[width=\textwidth]{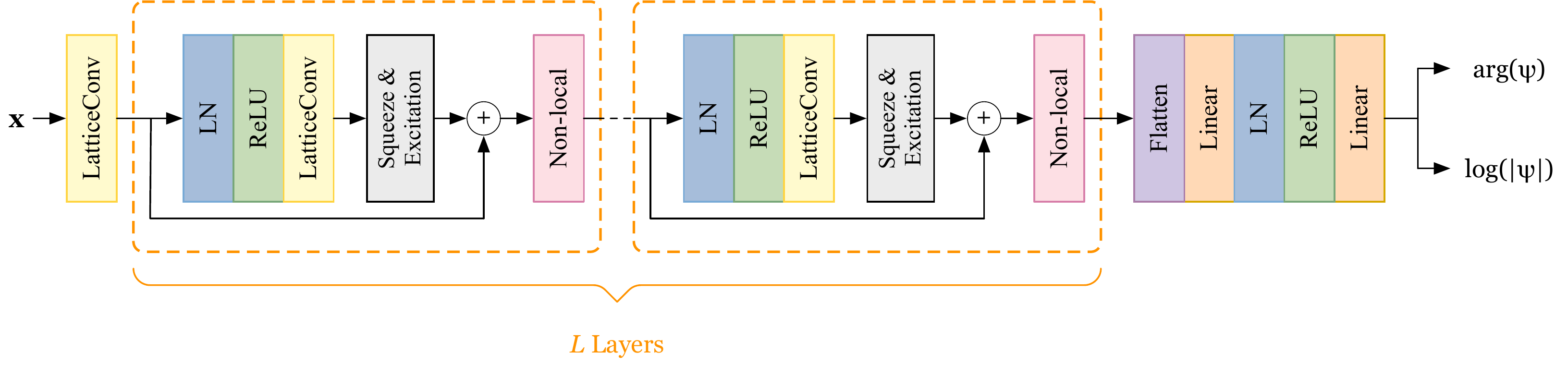}}
\caption{Lattice Convolutional Networks architecture}
\label{fig:cnn_architecture}
\end{center}
\vskip -0.2in
\end{figure*}

In our model, we predict amplitudes and phases separately with real-valued network, following~\citet{kochkov2021learning}. We first use lattice convolution to transform original spin configuration input into embedding space and then stack multiple SE-Non-Local layers to obtain the final latent representation. At last, we flatten the feature maps into a vector to keep all information on each lattice vertex and use MLP to obtain the log amplitude and argument of the wave function value. The overall architecture of our variational ansatz is shown in \cref{fig:cnn_architecture}.

\textbf{Training. }
As discussed in Section \ref{section:varional learning}, we use Variational Monte Carlo framework to optimize the variational ansatz iteratively by minimizing the statistic expectation of system energy $\langle E \rangle$. The data (spin configurations) are sampled from the probabilistic distribution defined by variational ansatz (neural network wave function) itself. Then we stochastically estimate the gradient $\nabla_w \langle E \rangle$~\cite{kochkov2021learning} to update network parameters $\mathbf{w}$. Details of the training method can be found in~\cref{app:Training Algorithm} and~\cref{app:VMC}.

\section{Experiments}
\label{sec:exp}

In this section, we evaluate the proposed LCN on learning ground states of the spin-1/2 $J_1 \text{-} J_2$ Heisenberg model, where the input vertices represent quantum spins. We show that our model can accurately approximate the ground state energies and achieve on par or better results with GNN models.


\textbf{Lattice.}
Following~\citet{kochkov2021learning}, four kinds of lattice are used in our experiments including square, honeycomb, triangular and kagome. Periodic boundary conditions are used so that the neighborhood patterns of the boundary vertices are the same with the internal vertices. All the lattice geometries we use in the experiments can be found in~\citet[Appendix A.5.a]{kochkov2021learning}.

\textbf{$\mathbf{J_1\text{-}J_2}$ Quantum Heisenberg model.}
The $J_1\text{-}J_2$ quantum Heisenberg model is the prototypical model for studying the magnetic properties of quantum materials. Its Hamiltonian matrix is given by:
\begin{align}
    \mathbf{H} = \sum_{\langle i,j \rangle} \mathbf{S}_i\cdot \mathbf{S}_j + J_2 \sum_{\langle\langle i,j \rangle\rangle} \mathbf{S}_i \cdot \mathbf{S}_j,
\end{align}
where $\mathbf{S}_i = (S^x_i, S^y_i,S^z_i)$ are the spin-1/2 operators of the $i$-th vertex. The spin operator $S_i^\alpha$ is an Hermitian matrix of dimension $2^N$, defined as 
$S^\alpha_i = I^{\otimes j-1}\otimes \sigma^\alpha \otimes I^{\otimes N-j}$, where $\otimes$ stands for Kronecker product, $I$ is the two-by-two identity matrix and $\sigma^\alpha$ is the two-by-two Pauli matrix for $\alpha=x,y,z$. The term $\mathbf{S}_i\cdot \mathbf{S}_j$ describes the antiferromagnetic exchange between the spin on site $i$ and the spin on site $j$. $\langle \cdot, \cdot \rangle$ denotes the nearest neighbors and $\langle\langle \cdot, \cdot \rangle\rangle$ denotes the second nearest neighbors, both in terms of euclidean distances; $J_2$ controls the interaction strength between the next nearest neighbors. The interaction strength between nearest neighboring spins is set to 1 as the unit.

\textbf{Setup. }
We test our models on four kinds of lattice with various sizes and $J_2$ values. We choose specific J2 values under which the ground state is much harder to learn for the GNN model. The \textbf{energy per site} of the learned wave function is used as evaluation metric where the total energy of the system is divided by the number of spins (vertices). A lower energy per site indicates a more accurate approximation to the ground state. 

We use the same CNN architecture for all lattices. The convolution operation in each CNN layer is replaced with the proposed lattice convolution according to the lattice type. Pre-activation is not used for triangular lattices. For kagome lattice of size 108, mask processing after convolution is not used. 

We implement our models as well as the VMC procedures in PyTorch~\cite{paszke2019pytorch} and all models were trained on NVIDIA RTX A6000. For efficiency we implement the VMC procedures with GPU parallelization. A batch of configurations are sampled or measured in parallel at each step. Samples are kept at intervals to minimize the correlation between consecutive samples where the interval is equal to the size of the system. We further initialize each MCMC chain with a different random seed to maximize the independence between chains. During training, we use the stochastic gradient estimated from each mini-batch of samples and optimize the models with the Adam optimizer~\cite{kingma2015adam}. The optimization generally converges within 30,000 steps. We save the model with the lowest stable training energy for testing, where the energies in neighboring steps have small variance. During testing, the energy is estimated from 200,000 configurations sampled from equilibrated Markov chains. Hyperparameters for optimization can be found in~\cref{app:hyper_params}. To estimate the stochastic error bars, we group Markov chains into 100 bins. The error bar is computed as the standard deviation of average energies from each bin.

\begin{table*}[t]
    \centering
    \caption{Estimated energy per site of the learned wave function (with error bars in parenthesis). Lower is better. Four kinds of lattice with various sizes are used for comparison. The $J_2$ value controls next nearest neighboring interaction on the lattice, resulting in differences in the ground states. The GNN-2 model doubles the parameters and computation by using separate branches for predicting the amplitude and argument parts of the wave function. Therefore, we directly compare our model against the single branch GNN model, and LCN results better than GNN are denoted in bold. We use $*$ to denote the results measured from plots in their reference papers. For the reference ground state energies we also annotate the employed methods: $\dagger$ for exact diagonalization, $\infty$ for infinite-size estimates, $\ddag$ for RMB+PP, $\S$ for QMC methods, and $\P$ for DMRG methods.}
    \vskip 0.1in
    \resizebox{0.9\textwidth}{!}{
    \begin{tabular}{l c c c c c c}
    \toprule
        Lattice & Size & $J_2$ & GNN & GNN-2 & LCN (ours) & Reference Energy \\
        \midrule
        \multirow{4}{*}{Square} & \multirow{2}{*}{36} & 0 & -0.6788$^*$ & - & -0.6788(1) &  -0.678872$^\dagger$ \\
        && 0.5 & -0.5022(4) & -0.5023(5) & -0.5022(2) & -0.503810$^{\dagger}$ \\\cmidrule{2-7}
        & \multirow{2}{*}{100} & 0 & -0.6708(0) & - & -0.6708(1) & 0.671549(4)$^\S$ \\
        && 0.5 & -0.4955(4) & -0.4960(5) & \textbf{-0.4957(4)} & -0.497629(1)$^\ddag$\\
        \midrule
        \multirow{4}{*}{Honeycomb} & \multirow{2}{*}{32} & 0 & -0.551$^*$ &-& -0.5516(1) & -0.5517$^{\dagger *}$\\
        && 0.2 & -0.4563(6) & -0.4564(7) & \textbf{-0.4563(1)} & -0.4567$^{\dagger *}$\\\cmidrule{2-7}
        & \multirow{2}{*}{98} & 0 & - & - & -0.5421(4) & -0.5448$^{\infty *}$\\
        && 0.2 & -0.4528(2) & -0.4536(5) & \textbf{-0.4538(4)} & -0.4527$^{\infty*}$\\
        \midrule
        \multirow{5}{*}{Triangular} & \multirow{3}{*}{36} & 0 & - & -0.55889 & \textbf{-0.5601(4)} & -0.5603734$^{\dagger}$\\
        && 0.08 & -0.5221$^*$ & - & \textbf{-0.5273(4)} & -0.5286$^{\dagger *}$\\
        && 0.125 & -0.512(2) & -0.5131(8) & \textbf{-0.5126(6)} & -0.515564$^{\dagger}$ \\\cmidrule{2-7}
        & \multirow{2}{*}{108} & 0 & -0.5508(8) & -0.5519(4) & -0.5475(4) & -0.551$^{\infty}$ \\
        && 0.125 & -0.500(9) & -0.5069(8) & \textbf{-0.5110(4)} & -0.5126$^{\P}$\\
        \midrule
        \multirow{3}{*}{Kagome} & \multirow{2}{*}{36} & 0 & -0.434(1) & -0.4338(6) & \textbf{-0.4367(4)} & -0.43837653$^{\dagger}$ \\
        && -0.02 & -0.4339$^*$ & - & \textbf{-0.4384(5)} & -0.4399$^{\dagger *}$\\\cmidrule{2-7}
        & 108 & 0 & -0.4302(1) & -0.4314(7) & -0.4276(3) & -0.4386$^{\P}$\\
        \bottomrule
    \end{tabular}}
    \vskip -0.2in
    \label{tab:result}
\end{table*}

\textbf{Results. }
The experimental results are summarized in \cref{tab:result}. We compare the proposed LCN with GNN~\cite{kochkov2021learning} as well as reference energies. We use the same references as in~\citet{kochkov2021learning}. 
For small systems ($N=32,36$), the reference energies are the exact ground state energies computed by direct diagonalizing the Hamiltonian matrix~\cite{schulz1996magnetic, albuquerque2011phase, iqbal2016spin, changlani2018macroscopically}. For large systems ($N=98,108$), the exact diagonalization is computationally infeasible. Reference energies are calculated with quantum Monte Carlo (QMC)~\cite{sandvik1997finite} and RBM+PP (pair-product states)~\cite{nomura2020dirac} for square with $J_2=0.0$ and $J_2=0.5$, respectively, and density-matrix renormalization group (DMRG)~\cite{iqbal2016spin, yan2011spin} for triangular and kagome. In some cases, the estimates for infinite-size lattice are used as reference energies, extrapolated from exact diagonalizations for honeycomb or from DMRG for triangular. 
Results show that the wave functions learned by the proposed LCN consistently gives energies close to the reference ground state energies on both small and large systems. For small systems, LCN gives similar ground state energies with GNN on square and honeycomb lattices while achieves better ground state energies on triangular and kagome lattices. For large systems at $J_2=0$, LCN achieves similar ground state energies with GNN on the square lattice while falls behind on triangular and kagome lattices. For large systems at $J_2\neq 0$, LCN  outperforms GNN on all tested lattices including square, honeycomb and triangular. The good performance proves that the proposed LCN is able to accurately represent the ground states of quantum many-body systems without explicit structure encoding. 

\textbf{Quantum State and Energy. }The quality of the variational result is measured by the ratio $(E-E_0)/\Delta$, where $\Delta$ is the gap of the Hamiltonian, i.e. the difference between the energies of the first excited state and the ground state. However the precise value of  gap $\Delta$ for most cases is not known  from the previous literature. The gap is typically an order one number in the unit of $J_1$ for a gapped system or a polynomial of $1/N$ for gapless systems. As a result, order one improvement on the total variational energy $E$, or order $1/N$ improvement on the energy per spin, listed in Table~\ref{tab:result}, is significant in approaching the true ground state. We conduct an experiment to show the relationship between quantum state and energy accuracy on 12 node kagome lattice in~\cref{app:state overlap}. 

\textbf{Model Capacity Comparison. }
We also compare the model capacity between LCN and GNN in~\cref{app:model_capacity}. On small lattices, our model with less trainable parameters than GNN can achieve performance on par or better than GNN. However, as described in~\cref{sec:network architecture}, we use flatten operation before final MLP layer to keep all information on lattices, so the number of our model parameters increases with system size. Nevertheless, experimental results on small systems demonstrate that our model is more expressive than GNN, and also characterize more accurate subspace of Hilbert space where wave functions reside in.  

\textbf{GNN with and without sublattice encoding. }
\begin{table}[t]
\parbox{.48\linewidth}{
    \centering
    \caption{Reproduced GNN performance with and without sublattice encoding.}
    \vskip 0.14in
    \resizebox{0.48\textwidth}{!}{
    \begin{tabular}{l@{\hspace{0.2cm}}c@{\hspace{0.2cm}}c >{\centering\arraybackslash}m{4.5em} >{\centering\arraybackslash}m{5.5em}}
        \toprule
        Lattice & Size & $J_2$ & GNN with sublattice & GNN without sublattice\\
        \midrule
        Square & 36 & 0.0 & -0.6776(6) & -0.6536(11)\\
        \midrule
        Honeycomb & 32 & 0.2 & -0.4551(3) & -0.4498(48)\\
        \midrule
        \multirow{3}{*}{Triangular} & \multirow{3}{*}{36} & 0.0 & -0.5571(5) & N/A\\
        && 0.08 & -0.5253(5) & N/A\\
        && 0.125 & -0.5119(6) & N/A\\
        \midrule
        Kagome & 36 & 0.0 & -0.4345(5) & N/A\\
        \bottomrule
    \end{tabular}}
    \vskip 0.11in
    \label{tab:sublattice}
}
\hfill
\parbox{.48\linewidth}{
    \centering
    \caption{Performance comparison between regular kernel and special kernel on honeycomb, triangular and kagome lattices.}
    \resizebox{0.48\textwidth}{!}{
    \begin{tabular}{l@{\hspace{0.2cm}}c@{\hspace{0.2cm}}c >{\centering\arraybackslash}m{4.5em} >{\centering\arraybackslash}m{4.5em}}
        \toprule
        Lattice & Size & $J_2$ & Regular Kernel & Special Kernel \\
        \midrule
        \multirow{2}{*}{Honeycomb} & \multirow{2}{*}{32} & 0 & -0.5516(1) & -0.5512(2)\\
        && 0.2 & -0.4563(1) & -0.4547(3)\\
        \midrule
        \multirow{3}{*}{Triangular} & \multirow{3}{*}{36} & 0 & -0.5601(4) & -0.5501(6)\\
        && 0.08 & -0.5273(4) & -0.5127(7)\\
        && 0.125 & -0.5126(6) & -0.4936(7)\\
        \midrule
        \multirow{2}{*}{Kagome} & \multirow{2}{*}{36} & 0 & -0.4367(4) & -0.4207(6)\\
        && -0.02 & -0.4384(5) & -0.4253(6)\\
        \bottomrule
    \end{tabular}}
    \label{tab:kernel_comparision}
}
\end{table}
To illustrate the necessarity of using hand-crafted sublattice encoding in the GNN model, we try to implement the GNN model based on details provided in~\citet{kochkov2021learning} and reproduce the results.
We conduct ablation study on sublattice encoding and summarize the results in \cref{tab:sublattice}. For triangular and kagome lattices, we cannot get valid results without using sublattice encoding (either too large or too small). One possible reason is that the distribution of learned wave function is very hard for accurate sampling. On square and honeycomb lattices, sublattice encoding is essential for achieving better results. So we can draw the conclusion that the hand-crafted sublattice encoding is important for GNN model. 

\textbf{Kernel Design Comparison. }
Apart from applying regular convolution kernels on augmented lattices, we also design special convolution kernels for original honeycomb, triangular and kagome lattices, which only capture the nearest neighbors of each center vertices. Details of these special kernel design can be found at~\cref{app:special kernel}.

We compare the performance of these two categories of kernel design on small systems over honeycomb, triangular and kagome lattices, as shown in \cref{tab:kernel_comparision}. Experimental results show that regular convolution kernels consistently outperform special kernels. We argue the reasons are two folds. First, the regular kernel directly captures interaction between part of second nearest neighbors, which is more helpful when $J_2$ is not zero. Second reason is that information reuse is hindered to some extent for special kernels. For example, when using special kernels for original kagome lattices, three distinctive kernel shapes need to be used for capturing different local structures on lattices. And the same local structure has little overlap with each other, resulting in less overlap between same kernels. This impedes the information reuse among the same local structures that is crucial for capturing long-range spin correlations~\cite{liang2018solving}. 


So we conclude that lattice augmentation together with regular kernels are necessary for processing quantum lattice systems. On one hand, lattice augmentation is general for these four kinds of lattice and does not need any prior domain knowledge compared with GNN's sublattice encoding. On the other hand, with virtual vertices added, regular kernels can be applied, which has many advantages over special kernels such as boosting information exchange and reuse, as described in~\cref{sec:Augmented lattice}.

\textbf{Different Kernels for Honeycomb. }
For honeycomb lattices, the special kernel has half less parameters than regular kernel. So for fairness comparison, we enlarge the special kernel to capture all next nearest neighbors and hence the number of parameters is similar to the regular kernel. As shown in~\cref{tab:honeycomb_2hop}, performance of the 2-hop special kernel is still worse than regular kernel with virtual vertices added. Another observation is that the 2-hop special kernel performs even slightly worse than 1-hop special kernel, which implies that enlarging the receptive field of special kernel cannot help capture further useful spin-spin correlation.

\begin{table}
    \centering
    \caption{Experiments on honeycomb with different kernel design.}
    \vskip 0.1in
    \resizebox{0.56\textwidth}{!}{
    \begin{tabular}{l@{\hspace{0.2cm}}c@{\hspace{0.2cm}}c >{\centering\arraybackslash}m{4.5em} >{\centering\arraybackslash}m{4.5em} >{\centering\arraybackslash}m{5.5em}}
        \toprule
        Lattice & Size & $J_2$ & Regular Kernel & Special Kernel & 2-hop Special Kernel\\
        \midrule
        Honey- & \multirow{2}{*}{32} & 0 & -0.5516(1) & -0.5512(2) & -0.5511(2)\\
        comb&& 0.2 & -0.4563(1) & -0.4547(3) & -0.4539(4)\\
        \bottomrule
    \end{tabular}}
    \vskip -0.1in
    \label{tab:honeycomb_2hop}
\end{table}

\section{Conclusion}
We propose lattice convolutions to process non-square lattices by converting them into grid-like augmented lattices through a set of operations. So regular convolution can be applied without using hand-crafted structure encoding, which is needed in the previous GNN method. And we design lattice convolutional networks that use self-gating and attention mechanisms to capture channel-wise interdependencies and spatial long-range spin correlations, which contribute to high expressivity of variational wave functions. We experimentally demonstrate the effectiveness of lattice convolution network wave functions and achieve performance on par or better than existing methods. 

\section*{Acknowledgments}
We thank Dmitrii Kochkov for answering our questions about their method and providing all the lattice data, and Meng Liu for valuable discussion on network design. This work was supported in part by National Science Foundation grant IIS-1908198 and National Institutes of Health grant U01AG070112.

\nocite{glasser2018neural,chen2018equivalence,yang2020deep,teng2018machine,han2019solving,luo2019backflow,sharir2020deep,stokes2020phases,hibat2020recurrent,roth2020iterative,astrakhantsev2021broken,ferrari2019neural,broecker2017machine}

\bibliography{neurips_2022}
\bibliographystyle{plainnat}

\newpage
\appendix

\include{Appendix}

\end{document}

%% file: Appendix.tex












\title{Appendix: Lattice Convolutional Networks for Learning Ground States of Quantum Many-Body Systems}

%



\maketitle

\appendix

\section{Summary of Test Lattices}
\label{app:lattice}
\begin{table}[ht]
    \centering
    \caption{Test lattices settings with different lattice types, system sizes and $J_2$ values.}
    \vspace{0.1in}
    \begin{tabular}{lcc}
    \toprule
        Lattice & Number of Nodes & $J_2$ \\
        \midrule
        \multirow{2}{*}{Square} & 36 & 0, 0.5\\
        & 100 & 0, 0.5 \\
        \midrule
        \multirow{2}{*}{Honeycomb} & 32 & 0, 0.2\\
        & 98 & 0, 0.2 \\ 
        \midrule
        \multirow{2}{*}{Triangular} & 36 & 0, 0.08, 0.125\\
        & 108 & 0, 0.125 \\ 
        \midrule
        \multirow{2}{*}{Kagome} & 36 & 0, -0.02\\
        & 108 & 0 \\
        \bottomrule
    \end{tabular}
    \label{tab:data}
\end{table}

\section{Training Algorithm}
\label{app:Training Algorithm}
\begin{algorithm}[H]
   \caption{Training Algorithm of Lattice Convolutional Networks}
   \label{alg:training algorithm}
\begin{algorithmic}[1]
    \STATE {\bfseries Input:} Lattice structure $\mathcal{L}$, number of spin sites $N$, Lattice convolution network $\psi_{\theta}$ with trainable parameter $\theta$, learning rate $\alpha$, Markov Chain batch size $B$, initial annealing step $s$, measure step $m$
    \STATE Set different random seeds for each Markov chain
    \STATE Randomly initialize spin configurations $\mathcal{C} \in \{+1, -1\}^{B \times N}$ with equal number of $+1$ and $-1$
    \STATE $\hat{\mathcal{C}}$ $\leftarrow$ \textbf{Metropolis–Hastings}($\psi_{\theta}$, $\mathcal{C}$, s)
    \REPEAT
    \STATE $E_{total} \leftarrow 0$
    \FOR{$i=1$ {\bfseries to} $B$}
        \STATE $E_{total} \leftarrow E_{total} + \sum_{j} H_{ij} \frac{\psi\left(c_{j}\right)}{\psi(c_{i})}$
    \ENDFOR
    \STATE $\hat{E_0} \leftarrow \frac{1}{B} E_{total}$
    \STATE $\theta \leftarrow \theta - \alpha \nabla \hat{E_0}$
    \STATE $\hat{\mathcal{C}}$ $\leftarrow$ \textbf{Metropolis–Hastings}($\psi_{\theta}$, $\hat{\mathcal{C}}$, m)
    \UNTIL{$\hat{E_0}$ is converged}
\end{algorithmic}
\end{algorithm}

\section{Variational Monte Carlo}
\label{app:VMC}
As discussed in Section \ref{section:varional learning}, we use Variational Monte Carlo approach to optimize the variational ansatz by minimizing the statistic expectation of system energy. The equation is shown as below:
\begin{equation}
    \begin{aligned}
    E =
    \frac{\langle\psi|\mathbf{H}|\psi\rangle}{\langle\psi | \psi\rangle}&=\frac{\sum_{i,j}\langle\psi | c_{i}\rangle\left\langle c_{i}|\mathbf{H}| c_{j} \right\rangle\left\langle c_{j} | \psi\right\rangle}{\langle\psi | \psi\rangle},\\
    &=\frac{\sum_{i}|\psi(c_{i})|^{2}\left(\sum_{j} H_{ij} \frac{\psi\left(c_{j}\right)}{\psi(c_{i})}\right)}{\sum_{i}|\psi(c_{i})|^{2}},\\
    &=E_{c_{i} \sim D}\left[\sum_{j} H_{ij} \frac{\psi\left(c_{j}\right)}{\psi(c_{i})}\right],
    \end{aligned}
\end{equation}
where the energy expectation is computed over the probability distribution  $D(c_{i}) = \frac{|\psi(c_{i})|^{2}}{\sum_{i}|\psi(c_{i})|^{2}}$, and $\sum_{j} H_{ij} \frac{\psi\left(c_{j}\right)}{\psi(c_{i})}$ is refereed to as local energy. Specifically, at each optimization step, we first use Markov-Chain Monte Carlo (MCMC) method to sample system configurations $c_{i}$ from target distribution, which is given by the neural network or called variational wave function. Then we can stochastically estimate the gradient $\nabla_w \langle E \rangle$~\cite{kochkov2021learning} to update network parameters $\mathbf{w}$. Meanwhile we can evaluate energy expectation in every optimization step by taking average of all the local energy associated with each sampled configuration. The matrix multiplication between $\mathbf{H}$ and $|\psi\rangle$ can be performed efficiently because of the sparseness of Hamiltonian, which is determined by the lattice topology and the specific quantum system. Given a quantum system of $N$ spin-1/2, the dimension of $\mathbf{H}$ is $2^N$, however the typical number of nonzero values in each row is only of order $N$.


\section{Special Kernel Design}
\label{app:special kernel}
We can design special convolution kernel structures based on different repetitive local patterns of different lattices, where each type of kernels only capture all the nearest neighbor nodes of the center vertices. 

\begin{figure}[ht]
\vskip 0.2in
\centering
\subfigure[Honeycomb]{
\label{Fig.sub.1}
\includegraphics[width=0.3\columnwidth]{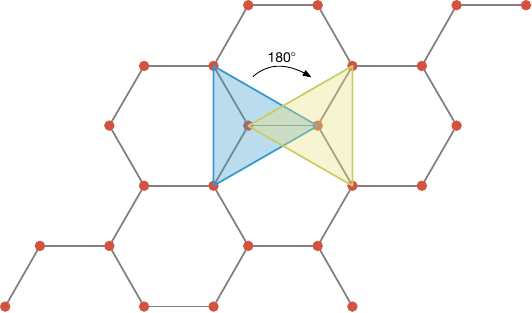}}\subfigure[Triangular]{
\label{Fig.sub.2}
\includegraphics[width=0.3\columnwidth]{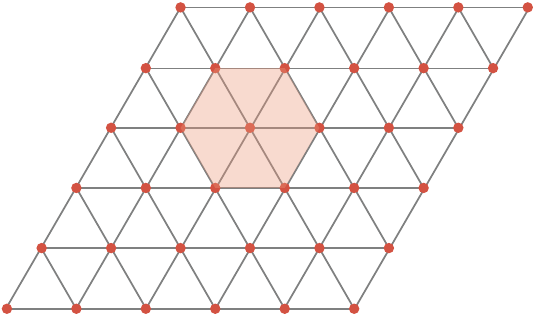}}\subfigure[Kagome]{
\label{Fig.sub.3}
\includegraphics[width=0.3\columnwidth]{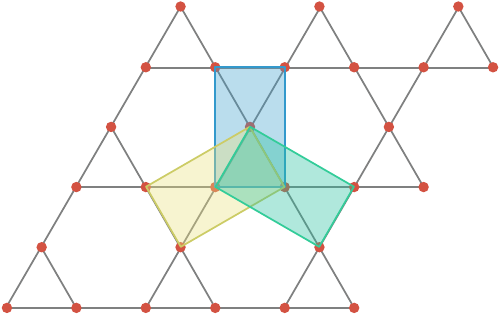}}
\caption{Special Kernel Design.}
\vskip -0.2in
\end{figure}

\paragraph{Honeycomb.}
We consider the convolution kernel that covers the nearest neighbors. For every vertex in the honeycomb lattice, its nearest neighbors always form an equilateral triangle. However, for two adjacent vertices, the orientations of the triangles formed by their neighbors are different. Specifically, two neighbors can be related by a 180-degree rotation. As a result, as shown in \cref{Fig.sub.1}, we rotate the convolution kernel by 180 degrees when going from one vertex to one of its neighbors.

\paragraph{Triangular.} For each center vertex, all of its nearest neighbors form a hexagon and this pattern is repetitive across the whole lattice. So we can naturally design hexagon shape convolution kernel to capture this pattern, as shown in \cref{Fig.sub.2}. In practice, we implement this kernel by masking two positions of the square kernel at the diagonal corner.

\paragraph{Kagome.} We observe that Kagome lattice has three different repetitive local patterns across the space, where each pattern can capture nearest neighbors around the center vertices in the pattern. So we design three different kernels corresponding to these three patterns, as shown in \cref{Fig.sub.3}. And these three kernels do not share weights with each other.

\section{Periodic Padding versus Zero Padding}
As discussed in ~\cref{sec:lattice_conv}, periodic padding is used before convolution in order to consider the periodic boundary condition for lattices. We compare the performance between periodic padding and zero padding, as shown in~\cref{tab:padding_comparision}, which shows that periodic padding is essential for achieving better results.
\begin{table}[ht]
    \centering
    \caption{Performance comparison between periodic padding and zero padding.}
    \vskip 0.15in
    \begin{tabular}{lcc >{\centering\arraybackslash}m{4.5em} >{\centering\arraybackslash}m{4.5em}}
        \toprule
        Lattice & Size & $J_2$ & Periodic padding & Zero padding \\
        \midrule
        \multirow{2}{*}{Honeycomb} & \multirow{2}{*}{32} & 0 & -0.5516(1) & -0.5511(2) \\
        && 0.2 & -0.4563(1) & -0.4542(3)\\
        \midrule
        \multirow{3}{*}{Triangular} & \multirow{3}{*}{36} & 0 & -0.5601(4) & -0.5109(6)\\
        && 0.08 & -0.5273(4) & -0.5095(5)\\
        && 0.125 & -0.5126(6) & -0.5091(5)\\
        \midrule
        \multirow{2}{*}{Kagome} & \multirow{2}{*}{36} & 0 & -0.4367(4) & -0.4356(5)\\
        && -0.02 & -0.4384(5) & -0.4349(4)\\
        \bottomrule
    \end{tabular}
    \vskip -0.1in
    \label{tab:padding_comparision}
\end{table}

\section{Quantum State Overlap and Energy}
\label{app:state overlap}
\begin{figure}[ht]
\vskip 0.2in
\begin{center}
\centerline{\includegraphics[width=0.85\columnwidth]{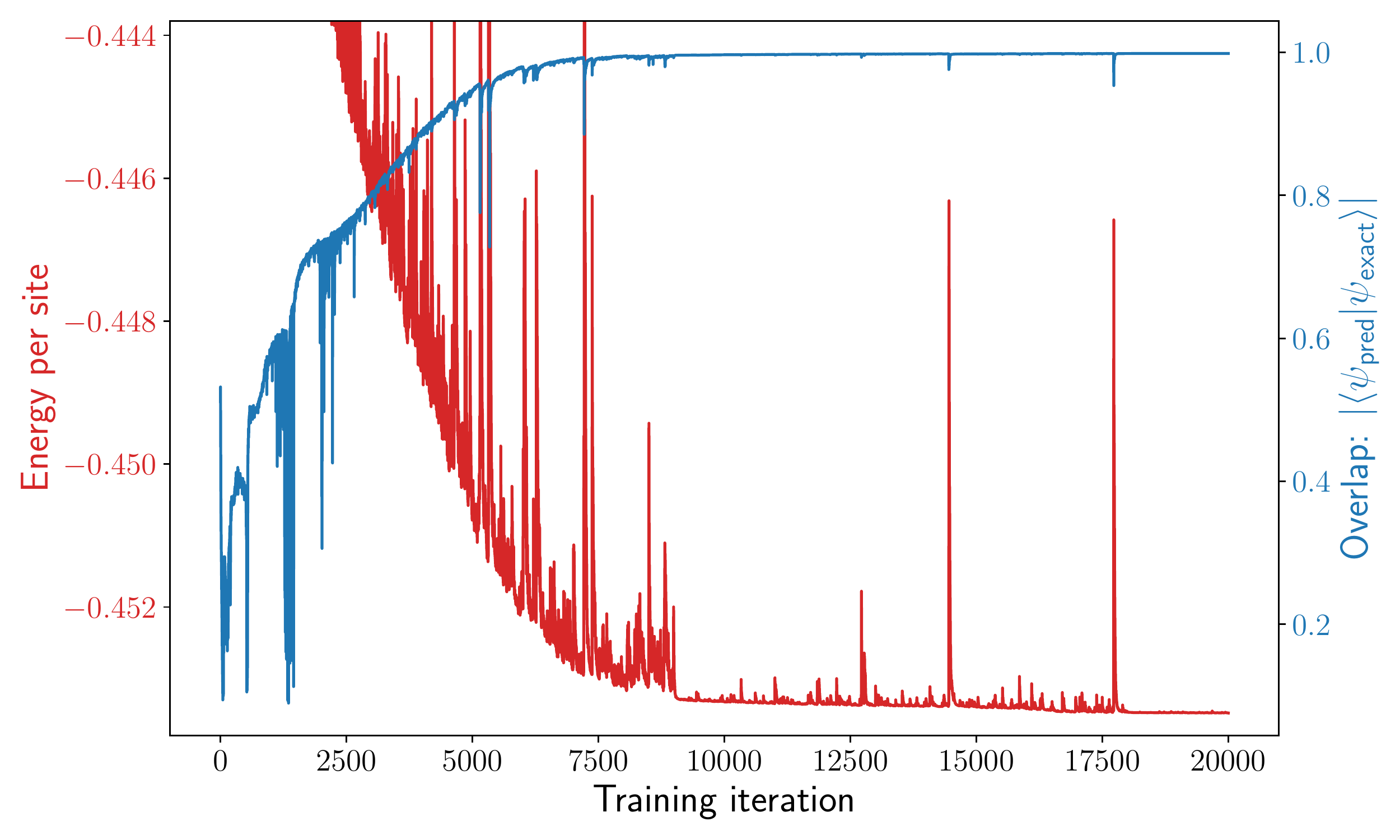}}
\caption{Improvement of state overlap on kagome lattice}
\label{fig:overlap}
\end{center}
\vskip -0.3in
\end{figure}

As described in main experimental results, our goal is to identify the ground-state. There could be dramatic improvements in approaching true ground-state even though the corresponding energy value improvement is small. We conduct an experiment on a small kagome lattice with 12 nodes to clarify this, where true ground state can be obtained. As shown in~\cref{fig:overlap}, we can observe that even though energy accuracy has 0.9\% improvement (from -0.4492 to -0.4533) , the ground state overlap (up to 1) can increase by 10\% (from 0.9087 to 0.9971).

\section{Model Capacity Comparison}
\label{app:model_capacity}
In this section, we compare the number of model parameters between Lattice Convolutional Networks (LCN) and GNN~\cite{kochkov2021learning} model over different lattices, as shown in  \cref{tab:param_comparision}. For triangular and kagome lattices with 36 nodes, LCN with only 0.29M and 0.49M trainable parameters is able to achieve better performance than the GNN model with 0.86M parameters. And on honeycomb and square lattices, LCN with 0.36M and 0.28M parameters yields same results with the GNN model. So we argue that LCN is more expressive and parameter efficient than GNN model on various small lattices. 

As for the lattices with larger nodes, the number of parameters in LCN is increased, which is caused by the final MLP layer and boundary alignment operation on the augmented lattices. 
Since we flatten the final feature maps into a vector which is then processed by a MLP layer, the number of parameters will increase with the size of the network input. Also, we need to zero-pad the augmented lattices into parallelograms due to the irregular boundaries of original lattice geometry structures, which then increases the size of the input feature map.

\begin{table}[ht]
    \centering
    \caption{Model parameters of Lattice Convolutional Networks and GNN. For square lattice of size 100, the number for LCN  is counted with 4 SE-Non-Local layers. We also use a 3-layer model which is slightly smaller and contains 1.0M parameters. See \cref{app:hyper_params} for details.}
    \vskip 0.1in
    \begin{tabular}{lc >{\centering\arraybackslash}m{5em} >{\centering\arraybackslash}m{4.5em}}
        \toprule
        Lattice & Size & LCN (Ours) & GNN \\
        \midrule
        \multirow{2}{*}{Square} & 36 & \textbf{0.28M} & 0.86M\\
        & 100 & 1.1M & 0.86M\\
        \midrule
        \multirow{2}{*}{Honeycomb} & 32 & \textbf{0.36M} & 0.86M\\
        & 98 &  1.9M & 0.86M\\
        \midrule
        \multirow{2}{*}{Triangular} & 36 & \textbf{0.29M} & 0.86M\\
        & 108 & 1.9M & 0.86M\\
        \midrule
        \multirow{2}{*}{Kagome} & 36 & \textbf{0.49M} & 0.86M\\
        & 108 & 4.6M & 0.86M\\
        \bottomrule
    \end{tabular}
    \vskip -0.1in
    \label{tab:param_comparision}
\end{table}


\section{Network Components}
\label{app:network}
\textbf{Squeeze-and-Excitation Block. }
The process of SE-block can be represented as~\cite{hu2018squeeze}:
\begin{equation}
    \begin{aligned}
    z_{c}&=\mathbf{F}_{s q}\left(\mathbf{u}_{c}\right)=\frac{1}{H \times W} \sum_{i=1}^{H} \sum_{j=1}^{W} u_{c}(i, j),\\
    \mathbf{s}&=\mathbf{F}_{e x}(\mathbf{z}, \mathbf{W})=\sigma(g(\mathbf{z}, \mathbf{W}))=\sigma\left(\mathbf{W}_{2} \delta\left(\mathbf{W}_{1} \mathbf{z}\right)\right),\\
    \tilde{\mathbf{x}}_{c}&=\mathbf{F}_{\text {scale }}\left(\mathbf{u}_{c}, s_{c}\right)=s_{c} \mathbf{u}_{c},
    \end{aligned}
\end{equation}

where $\mathbf{u}_{c} \in \mathbb{R}^{H \times W}$ denote the c-th output feature map of convolution operator. $z_{c}$ denote the c-th element of channel descriptor $\mathbf{z}$ squeezed from $\mathbf{u}_{c}$. $\delta$ refers to ReLU~\cite{nair2010rectified} activation function and $\sigma$ refers to sigmoid function. $\mathbf{W}_{1} \in \mathbb{R}^{\frac{C}{r} \times C}$ and $\mathbf{W}_{2} \in \mathbb{R}^{C \times \frac{C}{r}}$. $r$ is the dimension reduction factor. The final output $\tilde{\mathbf{x}}_{c}$ is computed by channel-wise multiplication between $s_{c}$ and $\mathbf{u}_{c}$.

\textbf{Non-Local Block. }
Response at a position, denoted as $\mathbf{y}_{i}$, is computed as a weighted sum of features at all positions. The output of each position is formulated as~\cite{wang2018non}:
\begin{equation}
    \begin{aligned}
        f\left(\mathbf{x}_{i}, \mathbf{x}_{j}\right)&=e^{\theta\left(\mathbf{x}_{i}\right)^{T} \phi\left(\mathbf{x}_{j}\right)},\\
    \mathbf{y}_{i}&=\frac{1}{\mathcal{C}(\mathbf{x})} \sum_{\forall j} f\left(\mathbf{x}_{i}, \mathbf{x}_{j}\right) g\left(\mathbf{x}_{j}\right), \\
    \mathbf{z}_{i}&=W_{z} \mathbf{y}_{i}+\mathbf{x}_{i},
    \end{aligned}
\end{equation}

where $g$ is a linear embedding that transform the input feature map $\mathbf{x}$: $g\left(\mathbf{x}_{i}\right)=W_{g} \mathbf{x}_{i}$. And  $f\left(\mathbf{x}_{i}, \mathbf{x}_{j}\right)$ is embedded gaussian. $\theta\left(\mathbf{x}_{i}\right)=W_{\theta} \mathbf{x}_{i}$ and $\phi\left(\mathbf{x}_{j}\right)=W_{\phi} \mathbf{x}_{j}$ are two embeddings to compute dot product similarity. $\mathcal{C}(\mathbf{x})=\sum_{\forall j} f\left(\mathbf{x}_{i}, \mathbf{x}_{j}\right)$ is normalization factor. $\mathbf{z}_{i}$ is the final output of this block at position $i$ by adding residual connection. 
\clearpage

\section{Hyperparameters}

\label{app:hyper_params}
\begin{table*}[hb]
    \centering
    \caption{Hyperparameters for optimization. The learning rate is multiplied by 0.1 at every decay step. During training, gradient norm is clipped from beginning for square and honeycomb lattices. For triangular and kagome lattices, gradient norm is clipped after first learning rate decay.}
    \vskip 0.15in
    \hspace*{-0.4in}
    \begin{tabular}{lcc >{\centering\arraybackslash}m{6em} c >{\centering\arraybackslash}m{4em} c >{\centering\arraybackslash}m{4em}}
        \toprule
        Lattice & Size & $J_2$ & Number of SE-Non-Local layers & Batch size & Learning rate & Learning rate decay steps & Gradient norm clip\\
        \midrule
        \multirow{4}{*}{Square} & \multirow{2}{*}{36} & 0 & 2 & 500 & 1e-3 & 20000, 40000, 60000 & 1\\
        && 0.5 & 2& 500 & 1e-3 & 20000, 40000, 60000 & 1 \\\cmidrule{2-8}
        & \multirow{2}{*}{100} & 0 & 3 & 200 & 5e-4 & 8000, 12000, 16000 & 1\\
        && 0.5 & 4 & 200 & 5e-4 & 8000, 12000, 16000 & 1\\
        \midrule
        \multirow{3}{*}{Honeycomb} & \multirow{2}{*}{32} & 0 & 2 & 500 & 1e-3 & 20000, 40000, 60000 & 1\\
        && 0.2 & 2 & 500 & 1e-3 & 20000, 40000, 60000 & 1\\\cmidrule{2-8}
        & \multirow{2}{*}{98} & 0 & 4 & 100 & 7e-4 & 8000, 12000, 16000 & 1\\
        && 0.2 & 4 & 100 & 7e-4 & 10000, 16000, 22000 & 1\\
        \midrule
        \multirow{5}{*}{Triangular} & \multirow{3}{*}{36} & 0 & 2 & 1000 & 1e-3 & every 4000 & 2\\
        && 0.08 & 2 & 1000 & 1e-3 & every 4000 & 2\\
        && 0.125 & 2 & 1000 & 1e-3 & every 4000 & 2\\\cmidrule{2-8}
        & \multirow{2}{*}{108} & 0 & 2 & 200 & 1e-3 & every 4000 & None\\
        && 0.125 & 2 & 200 & 1e-3 & every 4000 & None\\
        \midrule
        \multirow{3}{*}{Kagome} & \multirow{2}{*}{36} & 0 & 2 & 1000 & 1e-3 & every 4000 & 1\\
        && -0.02 & 2 & 1000 & 1e-3 & every 4000 & 1\\\cmidrule{2-8}
        & 108 & 0 & 2 & 200 & 1e-3 & every 4000 & 1\\
        \bottomrule
    \end{tabular}
    \vskip -0.1in
    \label{tab:hyper_param}
\end{table*}